\documentstyle[epsfig]{aipproc}

\def\siml{\lower4pt \hbox{$\buildrel < \over \sim$}}
\def\simg{\lower4pt \hbox{$\buildrel > \over \sim$}}
\def\Mesz{M\'esz\'aros~}

\def\etal{{\it et~al.}}
\def\bec{\begin{center}}
\def\enc{\end{center}}
\def\beq{\begin{equation}}
\def\enq{\end{equation}}
\def\bea{\begin{eqnarray}}
\def\ena{\end{eqnarray}}
\def\bsk{\bigskip}

\newcommand{\boxsize}{0.89\textwidth}

\begin{document}
{\footnotesize
\noindent Presented at the\hfill\\
{\it 5th Huntsville Gamma-Ray Burst Symposium,}\hfill\\
Hunstville, Oct. 18-22, 1999}
\bigskip

\title{ Photospheres, Comptonization and \\X-ray Lines in Gamma Ray Bursts }

\author{ P. \Mesz$^{1,2,3}$ } 

\address{$^1$Pennsylvania State University, 525 Davey, University Park, PA 16802 \\
$^2$California Institute of Technology, MS 105-24, Pasadena, CA 91125~~~~\\
$^3$E-mail address: nnp@astro.psu.edu} 


\maketitle

\begin{abstract}
Steep X-ray spectral slopes, X-ray excesses and preferred spectral energy breaks in 
the 01-0.3 MeV range are discussed as the possible consequences of the photospheric
component of the GRB relativistic outflow, and of pair breakdown in internal shocks
leading to comptonization on semirelativistic electrons or MHD waves. 
We also discuss the X-ray and UV spectra of GRB afterglows occuring in a
dense environment characteristic of massive stellar progenitors, including their
ability to produce detectable Fe or other metallic line features.

\end{abstract}

\maketitle

\section{Photospheres, Shocks and Pairs}
\label{sec:phot}
 
A significant fraction of bursts appear to have
low energy spectral slopes steeper than 1/3 in energy\cite{preece+98,crider+97}.
This has motivated consideration of a thermal or nonthermal\cite{liang+97,liang+99}
comptonization mechanism, while leaving the astrophysical model largely unspecified.
There is also evidence that the apparent clustering of the break energy
of GRB spectra in the 50-500 keV range may not be due to observational selection
\cite{preece+98,brainerd+98peak,dermer+99apjl}. 
Models using Compton attenuation \cite{brainerd+98apj}  
require reprocessing by an external medium whose column density adjusts itself to a 
few g cm$^{-2}$.  More recently a preferred break has been attributed to a blackbody
peak at the comoving pair recombination temperature in the fireball photosphere
\cite{eichlerlev99}.  
In order for such photospheres to occur at the pair recombination temperature in the 
accelerating regime requires an extremely low baryon load. For very large baryon 
loads, a different explanation has been invoked \cite{tho94}
involving scattering of photospheric photons off MHD waves in the photosphere,
which upscatters the adiabatically cooled photons up to the observed break energy. 

\begin{figure}[ht]
\begin{center}
\epsfbox{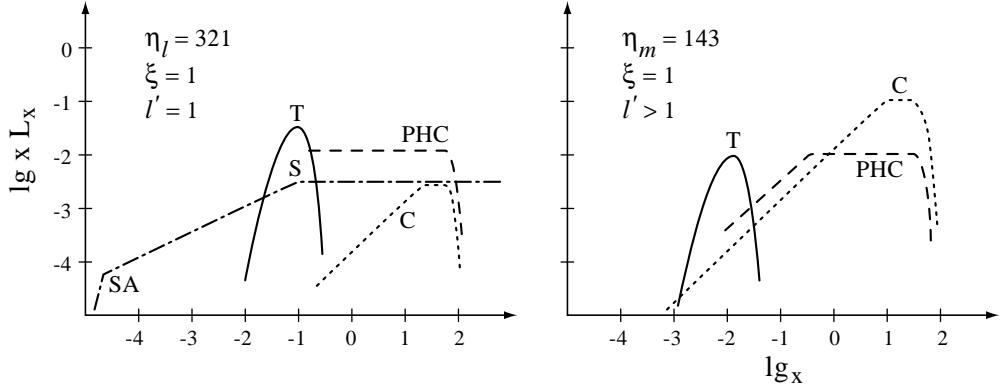}
\caption{\label{fig:photsp}  Luminosity per decade $xL_x$ vs. $x=h\nu/m_e c^2$ for two values of $\eta=L/{\dot M}c^2$ and marginal (left) or large (rigt) pair compactness.  T: thermal photosphere, PHC: photospheric comptonized component; S: shock synchrotron; C: shock pair dominated comptonized component (\Mesz \& Rees, 1999b).}
\end{center}
\end{figure}

Motivated by the above observations, these ideas have been synthesized \cite{mr99b}
to produce a generic scenario in which the presence
of a photospheric component as well as shocks subject to pair breakdown can produce
steep low energy spectra and preferred breaks (see Figure \ref{fig:photsp}).
In some of our previous work \cite{mlr93,rm94} considering photospheres and pair formation, 
their thermal character, the uncompensated photosphere redshift in the coasting phase, and the 
requirement of a power law extending to GeV energies were arguments in favor of a synchrotron and 
inverse Compton mechanism in shocks. The latter should, indeed, play a significant role in any 
model.  However, a photosphere is always present, even if not always dominant. If the
photosphere occurs in the accelerating regime where $\Gamma \propto r$, its energy is comparable 
to that of shocks which may occur further out, and the energy at which the black-body peak (T) 
is observed is in the ``magic" range near 0.5 MeV, for $\eta \ge \eta_\ast$, where
$\eta=L/{\dot M} c^2 \rightarrow \Gamma_f$ is the terminal bulk Lorentz factor and $\eta_\ast=
(L \sigma_T /4\pi m_p c^3 r_o )^{1/4} \simeq 10^3 (L_{52} r_7^{-1})^{1/4}$. Both its peak energy 
and its total energy are lower if the photosphere occurs in the coasting phase ($\eta \leq 
\eta_\ast$).  A steep low energy spectral slope is provided by the Rayleigh-Jeans part 
of the photosphere, and a low-energy excess or terrace by its Wien part. A high energy power 
law extending above this up to GeV requires, however, a separate explanation. One possibility 
is up-scattering of photospheric photons in the $\tau_T \simg 1$ region by Alfv\'en waves, 
whose energy may be a fraction of the bulk kinetic energy \cite{tho94}. This leads to a 
comptonized broken power law spectrum (PHC) in $xF_x$ ($x=h\nu/m_e c^2$) of slope 1 
up to the ``magic" break energy $x\siml 1$, and slope 0 up to $x\siml \eta$ above that
(Fig. \ref{fig:photsp}). The energy in this PHC wave-comptonized component can be substantial 
relative to the photosphere, and equals the ratio of wave to bulk kinetic energy. 

Above the photosphere, internal shocks are expected to occur \cite{rm94}, which would lead to 
a nonthermal Synchrotron/IC spectrum  (S) additional to the above. However, if the compactness 
parameter $\ell'$ (or comoving luminosity) is high, pair formation occurs 
which could produce a self-regulated low pair (comoving) temperature $\Theta'_p = kT'/m_e c^2 
\sim 10^{-1}$ favoring comptonization \cite{ghiscel99apjl}. In this $\ell' \geq 1$ case, thermal 
comptonization on the subrelativistic electrons leads to another comptonized component (C) of 
slope 1 up to an observer-frame energy $x\sim \Theta'_p \eta \sim 10^{-1}\eta$. Above this,
if scattering off waves also occurs in the shocks, a second component of slope 0 would
extend above it to $x \siml \eta$.

\section{X-ray and UV Line Spectra  of GRB}
\label{sec:lines}
 
The environment in which a GRB occurs may also lead, in the afterglow phase, to
specific spectral signatures from the external medium imprinted in the 
continuum, such as atomic edges and lines \cite{bkt97,pl98,mr98b}. These may be
used both to diagnose the chemical abundances and the ionization state (or local
separation from the burst), as well as serving as potential alternative redshift
indicators. (In addition, the outflowing ejecta itself may also contribute
blue-shifted edge and line features, especially if metal-rich blobs or filaments are
entrained in the flow from the disrupted progenitor debris \cite{mr98a}, which
could serve as diagnostic for the progenitor composition and outflow Lorentz factor).
An interesting prediction \cite{mr98b} is that an Fe {K-$\alpha$} X-ray {\it emission} line 
could be a diagnostic of a hypernova, since in this case one may expect a 
\begin{figure}[ht]
\begin{center}
\vspace*{-0.5cm}
\begin{minipage}[t]{0.6\textwidth}
\epsfxsize=\boxsize
\epsfbox{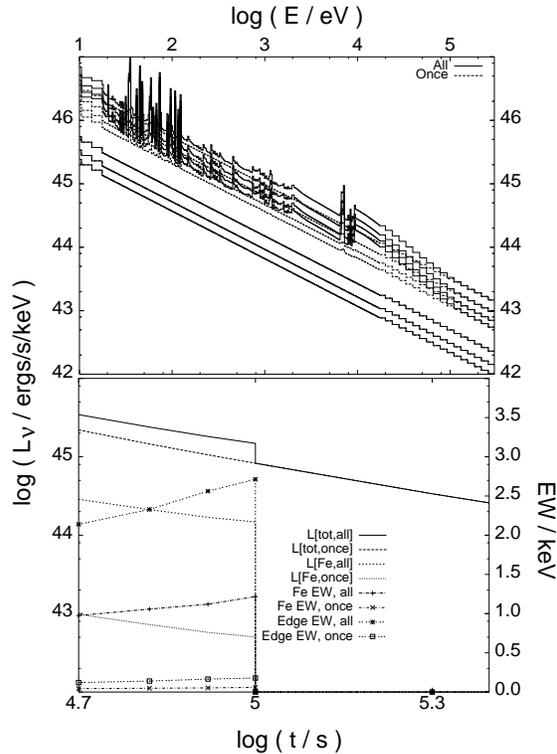}
\end{minipage}
\hspace{5mm}
\begin{minipage}[t]{0.32\textwidth}
\vspace*{-9cm}
\caption{\label{fig:funsp} Model hypernova funnel spectrum (top) for observer times of
50, 66, 83, 100, 200, 300 ks top to bottom, and (bottom) the total and Fe light curves 
and equivalent widths (Weth, \Mesz, Kallman \& Rees 1999), for with $R=1.5\times 10^{16}$ cm, 
$n=10^{10}$ cm$^{-3}$, and Fe abundance $10^2$ times solar.}
\end{minipage}
\vspace*{-0.5cm}
\end{center}
\end{figure}
massive envelope at a radius comparable to a light-day where $\tau_T \siml 1$, capable of 
reprocessing the X-ray continuum by recombination and fluorescence (see also 
\cite{ghi98,bot98}).  Detailed radiative transfer calculations have
been performed to simulate the time-dependent X/UV line spectra of massive
progenitor (hypernova) remnants\cite{weth+99}, see Figure \ref{fig:funsp}.
Two types of hypernova environment geometries were considered, which are illuminated
by a typical time-dependent broken power law afterglow continuum spectrum. One model
consists of a dense shell, such as a supernova remnant, which could be the product of an
inhomogeneous wind of variable velocity. This is essentially a transmission model, and 
produces initially an absorption X-ray line spectrum, turning later into an emmision spectrum, 
in which for Fe abundances 10 or 100 times solar the Fe line luminosities are $\siml 
10^{42}-10^{43}$ erg s$^{-1}$. 
The other model assumes a funnel geometry and is essentially a reflection model, 
with an empty or low density region along an axis, such as would arise in a rotating stellar 
envelope or a wind. The fireball and the afterglow propagate inside this funnel, which acts 
as a channel that collimates and reflects the continuum. This results in an emission line 
spectrum (Fig. \ref{fig:funsp}), where for 10 or 100 times solar abundances the Fe K-$\alpha$
line luminosity reaches $L_{Fe} \siml 10^{44}$ erg s$^{-1}$, with line and edge equivalent widths 
$EW\siml 1$ keV. This is comparable to the $3\sigma$ Fe features reported by two groups 
\cite{piro98b,yosh98} in GRB 970508 and GRB 970828. 

It is interesting that the Fe K-edge is significant in a funnel model such as shown
Fig \ref{fig:funsp}. While the energy of an Fe line 6.7 keV feature in GRB 970508 agrees with its 
previously known redshift $z=0.835$, the line feature of GRB 970828 would be in agreement with
the 9.28 keV Fe K-edge energy at this object's newly reported \cite{djorg+00} redshift of 
$z=0.958$. The line features in the 30-40 eV source-frame range seen in Figure \ref{fig:funsp} 
would be redshifted into the optical for $z \simg 5$, but are likely to be blanketed by the 
Ly-$\alpha$ forest of intervening high  redshift galaxies. However, it may be possible to detect
the soft X-ray metallic lines which become prominent soon after the Fe features, as the continuum 
softens and the gas cools, e.g. S and Si in the 2-3 keV source-frame range, or $\sim 1-1.5$ keV 
at $z\sim 1$. 

\bsk\noindent
I am grateful to M.J. Rees, C. Weth and T. Kallman for stimulating collaborations, 
NASA NAG-5 2857, the Guggenheim Foundation and the Division of Physics, Math \& Astronomy, 
the Astronomy Visitor and the Merle Kingsley funds at Caltech for support.


\begin{references}

\bibitem{bkt97} Bisnovatyi-Kogan, G \& Timokhin, A, 1997, Astr. Rep. 41, 423
\bibitem{bot98} B\"ottcher, M, \etal, 1998, astro-ph/9809156
\bibitem{brainerd+98apj} {Brainerd}, J. \etal (1998), ApJ, 501:325.
\bibitem{brainerd+98peak} Brainerd, J \etal (1999) in {\it Abstr 19th Texas Symp}, Paris (astro-ph/9904039).
\bibitem{crider+97} {Crider}, A. \etal (1997), ApJ, 479:L39.
\bibitem{dermer+99apjl} Dermer, C.D., \etal, 1999, Ap.J., 515, L49.
\bibitem{djorg+00} Djorgovski, S.G., \etal, 2000, Ap.J. subm
\bibitem{eichlerlev99} Eichler, D \& Levinson, A (1999), ApJ, subm(astro-ph/9903103).
\bibitem{ghi98} Ghisellini, G, \etal, 1998, astro-ph/9808156
\bibitem{ghiscel99apjl} {Ghisellini}, G. and {Celotti}, A. (1999), ApJ, 511, L93.
\bibitem{liang+97}  Liang, E \etal (1997), Ap.J., 491, L15.
\bibitem{liang+99} Liang, E. \etal (1999), Ap.J., 519, L21.
\bibitem{mlr93} \Mesz, P., Laguna, P \& Rees, M.J., 1993, ApJ, 415, 181
\bibitem{mr99b} \Mesz, P. \& Rees, M.J., ApJ, in press (1999) (astro-ph/9908126). 
\bibitem{mr98a} \Mesz, P \& Rees, M.J., 1998a, ApJ, 502, L105
\bibitem{mr98b} \Mesz, P \& Rees, M.J., 1998b, MNRAS, 299, L10
\bibitem{pl98} Perna, R. \& Loeb, A., 1998, ApJ, 503, L135
\bibitem{piro98b} Piro, L, \etal, 1998b, A\& A Supp.,in press (Procs. Rome Conference on GRB)
\bibitem{preece+98} {Preece}, R. \etal (1998), ApJ, 496, 849.
\bibitem{rm94} Rees, M.J. \& \Mesz, P., 1994, ApJ, 430, L93
\bibitem{tho94} Thompson, C., 1994, MNRAS, 270, 480
\bibitem{weth+99} Weth, C, \Mesz, P, Kallman, T \& Rees, M.J, ApJ 1999 (astro-ph/9908243)
\bibitem{yosh98} Yoshida, A, \etal, 1998, A\& A Supp.,in press (Procs. Rome Conf. on GRB)

\end{references}
\end{document}